# THz-Raman Identification of labile products in the system «phenol-semiquinone-quinone».

*E.A. Iasenko [1], Dr., V.P. Chelibanov[1], Dr., A.M. Polubotko[2]*
*[1]State University of Information Technologies, Mechanics, and Optics, St. Petersburg, 197101 Russia*
*[2]A.F. Ioffe Physico-Technical Institute, Russian Academy of Sciences, St. Petersburg, 194021 Russia*

***Abstract***

The paper presents the results of SERS studies of the dynamic behavior of "phenol-semiquinone-quinone" system. This system is a key part of chemiluminescent sensors for reactive oxygen species. The dynamics of the system seems to be very important in the processes that determine the secondary metabolism at the cellular level in molecular biology. THz Raman spectra were recorded for the labile products formed in the processes initiated by proton transfer. A mechanism of the proton-transfer-initiated reaction is proposed.

**Keywords - Benzoquinone, THz Raman spectrometer, 3,4,5-trihydroxybenzoic acid, kinetic.**

## Introduction

Redox-active organic compounds having several stable redox forms are of utmost interest for researchers creating highly-sensitive sensors, optical switching devices, catalyst systems, and high-efficiency pharmaceuticals, [1,2]. Among these, a special place is occupied by redox-active ligands capable of changing the degree of oxidation and thereby forming chelate structures with metal ions. This significantly enhances the redox properties of metal complexes (ability to raise the number of stable redox-transitions in the complexes) and improves the operation selectivity of catalyst systems. It is believed that triple redox-systems based on o-benzoquinone can usually exist in one of three valence forms: neutral, i.e., o-quinone form; anion-radical form (one-electron reduced anion-radical form); and catecholate-dianion form (two-electron reduced or deprotonated form of pyrocatechols). Complexes of transition metals are usually formed in the neutral o-quinone, o-semiquinone, and catecholate forms. A study of the process of partial or full protonation of o-quinones and their reduction to the initial phenol in adsorbed state, as well as identification of the type of a quinone being formed (ortho- or para-), is a vital task.

The experiments described in this paper were performed in order to study the photochromic behavior of the “quinone-semiquinone-phenol” redox system. Just this system is a key element of a heterogeneous chemiluminescent sensor for trace concentrations of ozone. In these experiments, the type of a quinone was identified by the spectral method and its protonation as a result of the photo-induced process, ESIPT (excited state intramolecular proton transfer), was examined. Hydroxy-benzoic acid in which hydroxyl groups replace hydrogen in positions 3, 4, and 5 positions (gallic acid) was chosen as a molecular object of study. Tetravalent titanium (in the form of titanium dioxide, a nanosize powder Degussa P-25, a mixture of minerals: 75% anatase and 25% rutile) served as a substrate with active centers for accommodation of ligands. The choice of $TiO_2$ as the substrate was predetermined by the polar properties of its surface, which provides stabilization of labile particles, and also by its ability to be studied by SERS (surface-enhanced Raman spectroscopy ). Owing to its spatial structure, gallic acid must mostly form quinones in the ortho-position in oxidation of terminal hydroxy groups (deprotonation). The oxidation selectivity of hydroxy groups increases when acid dimers are involved in the oxidation reaction. In fact, a sterically hindered polyphenol is formed in dimerization of gallic acid monomers at carboxy groups.

*A. Experimental section*

A 1M solution of gallic acid in ethanol was micro-pipetted onto the surface of compacted $TiO_2$ nanoparticles. The probable formation of a protonated structure of the gallic acid dimer on the substrate surface is shown in fig. 1. Then, the specimen was placed in the focal plane of a microscope, a confocal THz Raman spectrometer. The THz Raman spectrometer is shown schematically in fig. 2. In the experiment, the emission of a single-mode laser (wavelength 785 nm, power 25 mW, laser linewidth 100 MHz) was used for photo-induced protonation of terminal hydroxyl groups of the gallic acid dimer.

It is known that the process of ortho- and para-quinone protonation can be initiated by transfer of a superoxide electron ($O_2^{-}$) to form a semiquinone radical [3]. The superoxide itself is easily formed on the surface of nanosize $TiO_2$ particles, which form heterostructures of conjugated type with catecholate types of ligands. Hydroxyl (OH•) and $HO_2$• radicals are also present on the $TiO_2$ surface [4]. The process of superoxide formation under visible or near-IR irradiation is illustrated by fig. 3.

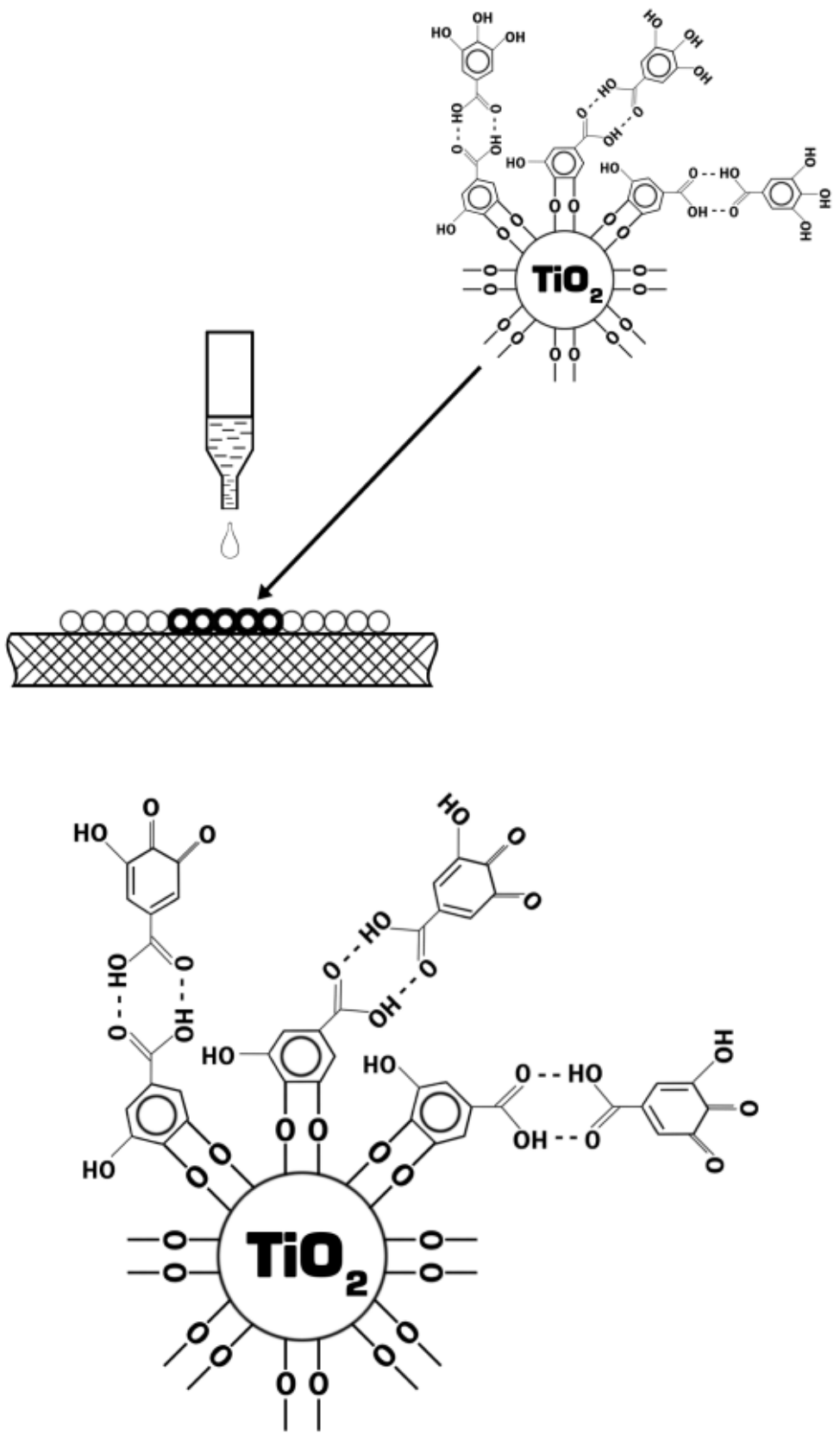

Fig.1. Probable formation of a protonated structure of the gallic acid dimer on the substrate surface.

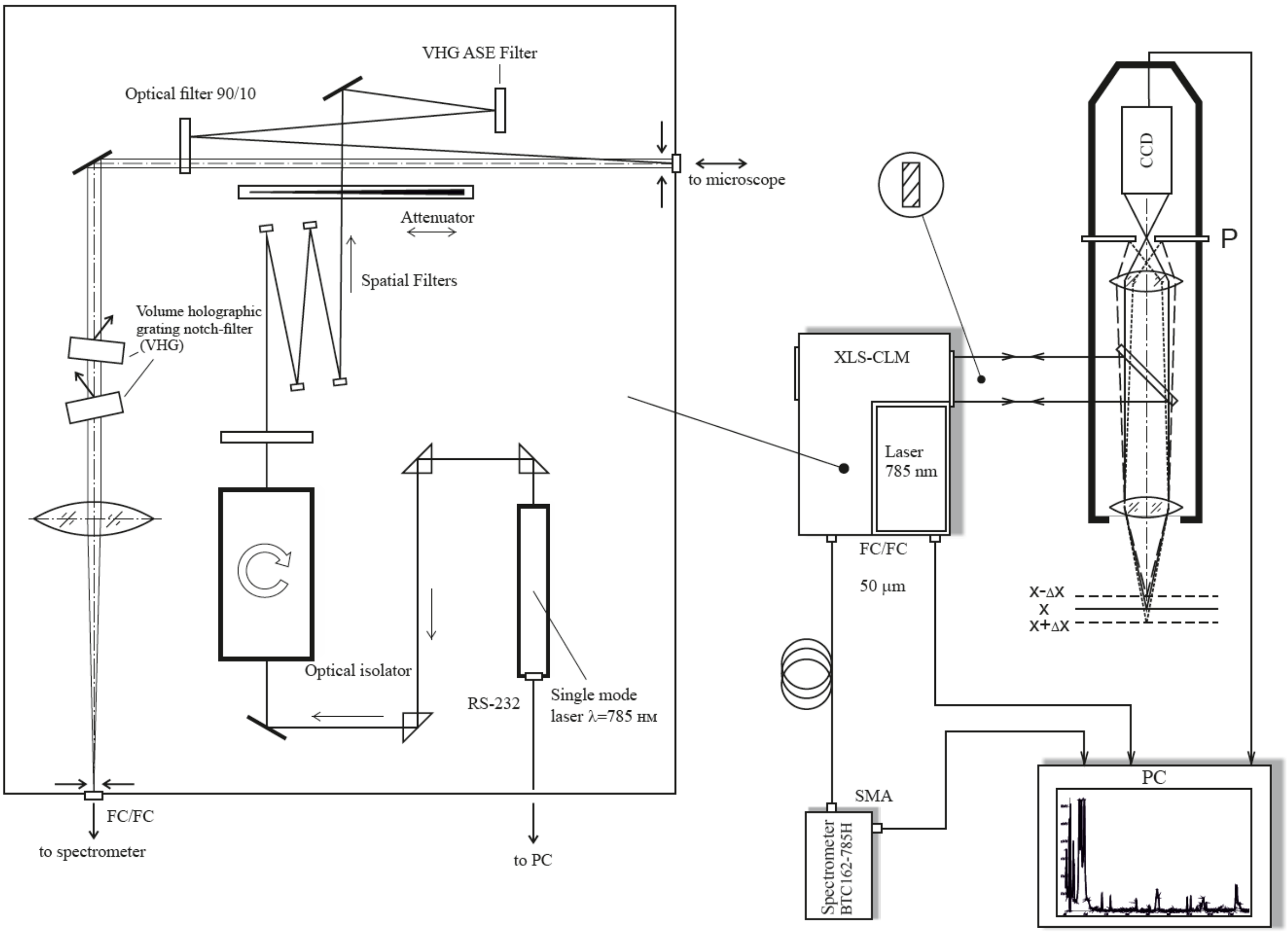

Fig.2. Block-diagram of the confocal THz Raman spectrometer for kinetic studies. The range of accessible Raman frequencies is from -49 to +1800 cm-1. X designates the focal plane of the microscope.

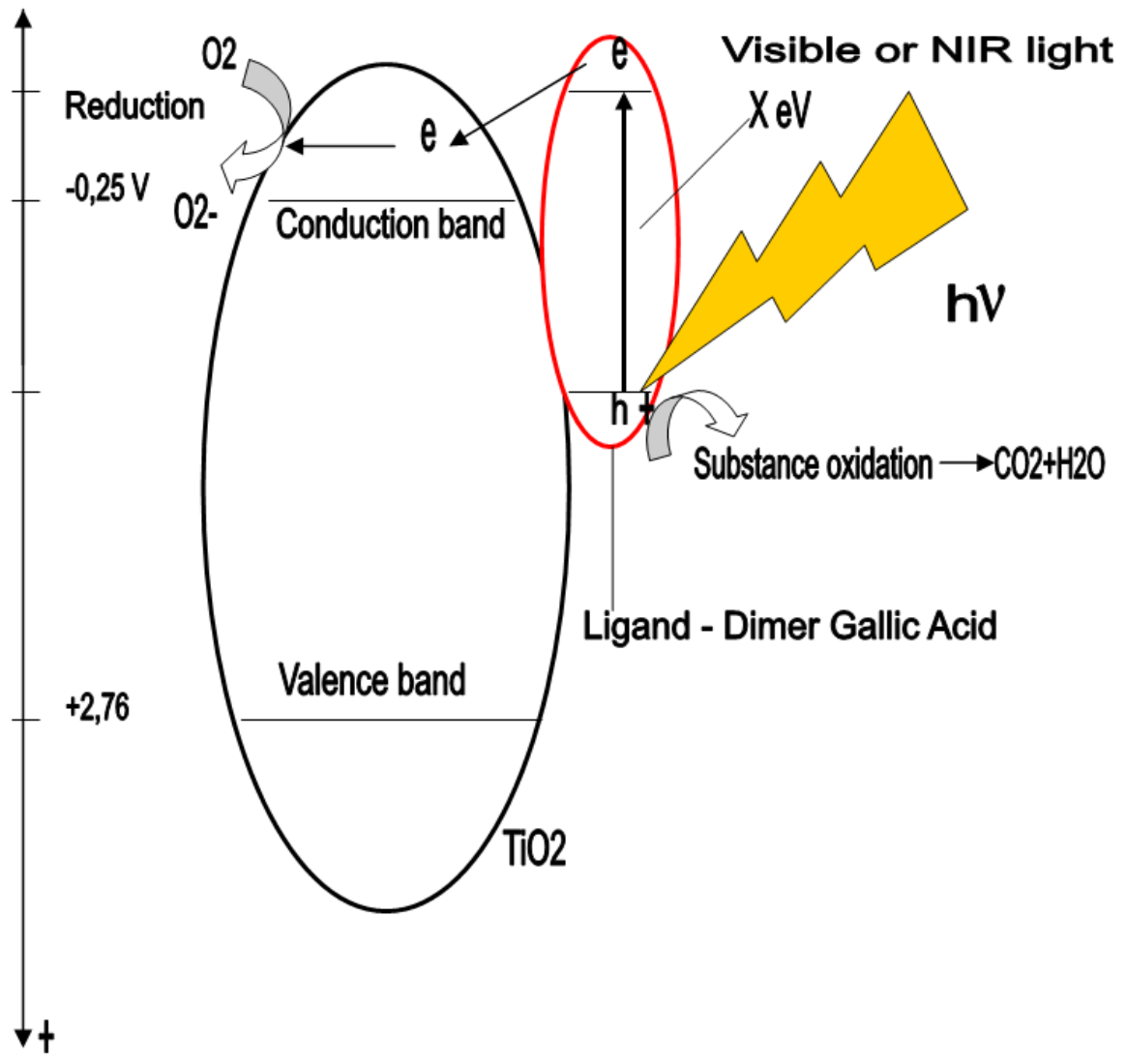

Fig.3. Formation of the superoxide in the reduction of oxygen by delocalized electrons from the conduction band of titanium dioxide. As the ligand of the catecholate type serves 3,4,5-trihydroxybenzoic (gallic) acid.

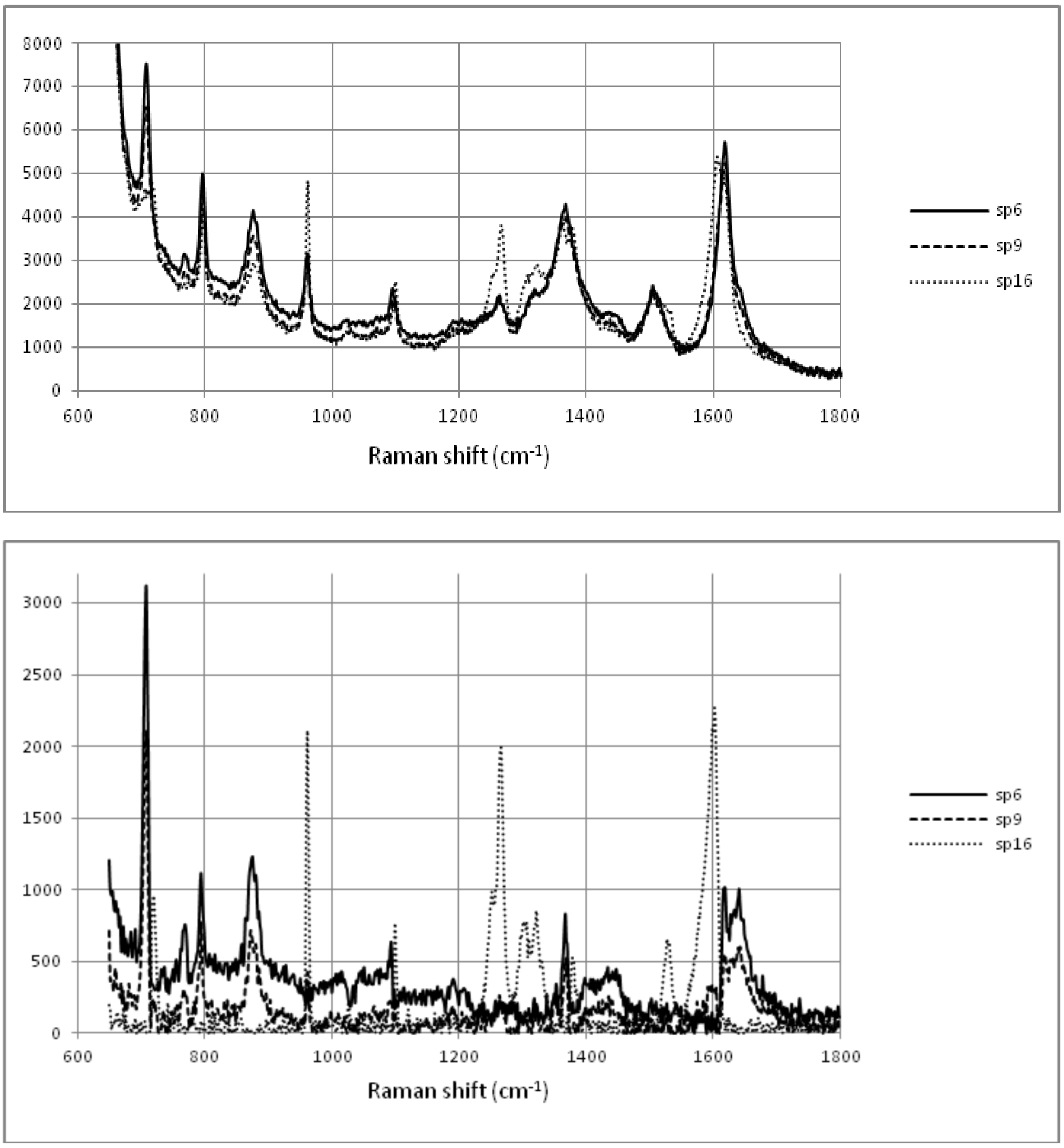

Fig. 4. (A) Raman spectra (for the range of 600…1800 cm-1) of the starting product, o-benzoquinone (SP6); final product of protonation, gallic acid (SP16); and probable product of one-electron reduction of quinone, semiquinone (SP9). (B) Raman spectra of o-benzoquinone, gallic acid, and semiquinone (after the deduction of the "minimal surface").

The minimal surface is a 2D surface, all points of which remain unchanged during the process. The minimal surface is deducted from the Raman signal surface obtained in the experiment. Of greater importance in this case are those points of the function that are the most strongly changed during the chemical reaction. Thus, the isolation and identification of individual substances in a multi-component mixture becomes possible due to their time separation during the process.

It was found that, during the photo-induced reduction of gallic acid from o-benzoquinone, the SERS spectrum of gallic acid originally adsorbed (in form of quinone) on the surface of titanium dioxide becomes similar to the spectrum of gallic acid in the free state (see fig. 7). This enables identification of the SERS spectrum of gallic acid (as a representative of catecholate ligands) by standard spectral databases.

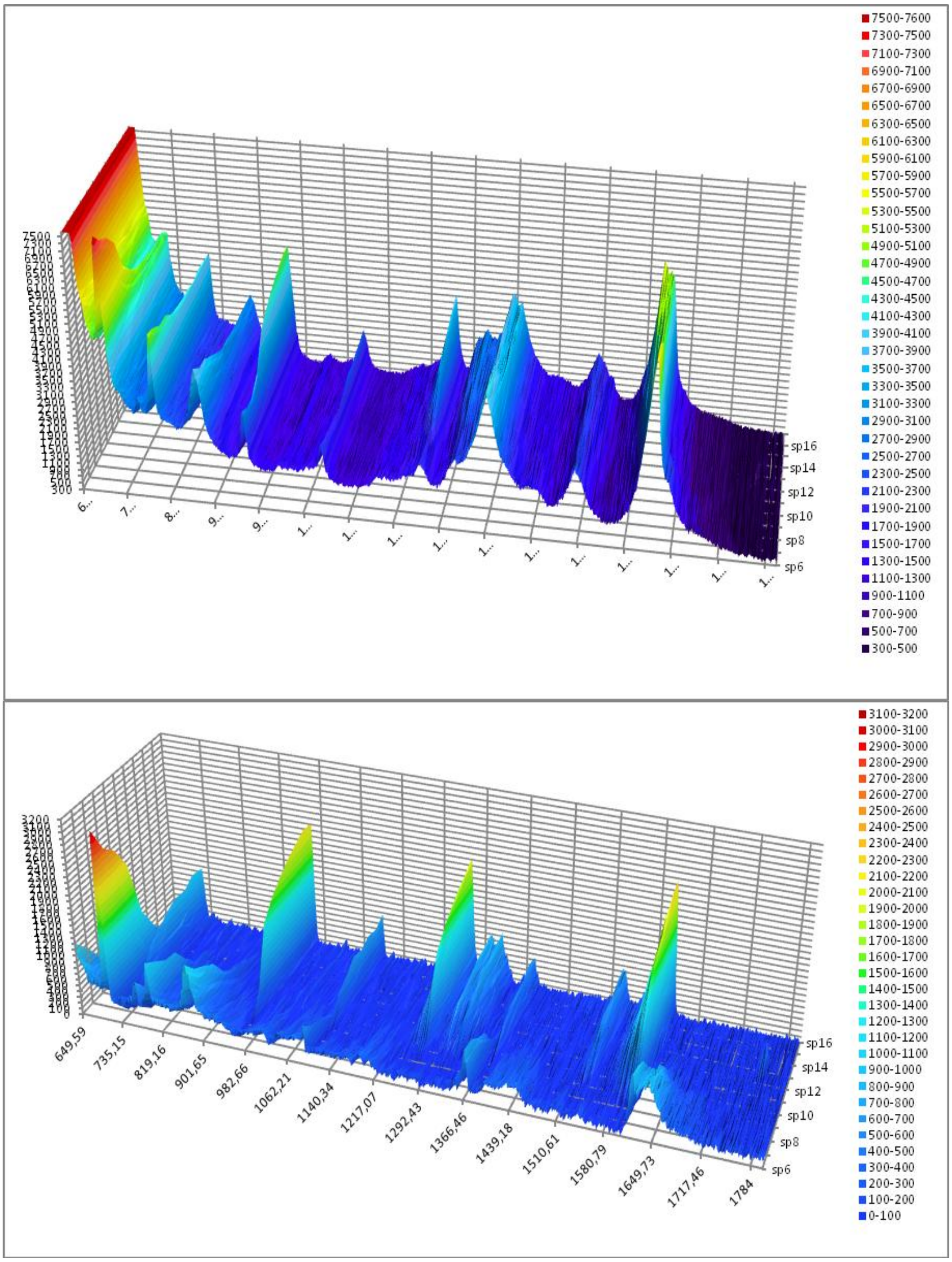

Fig.5. Kinetics of o-benzoquinone consumption and accumulation of gallic acid in the system during the photo-induced reduction of the product. (a) Reference Raman spectra, (b) the same spectra after the processing (deduction of the “minimal surface”).

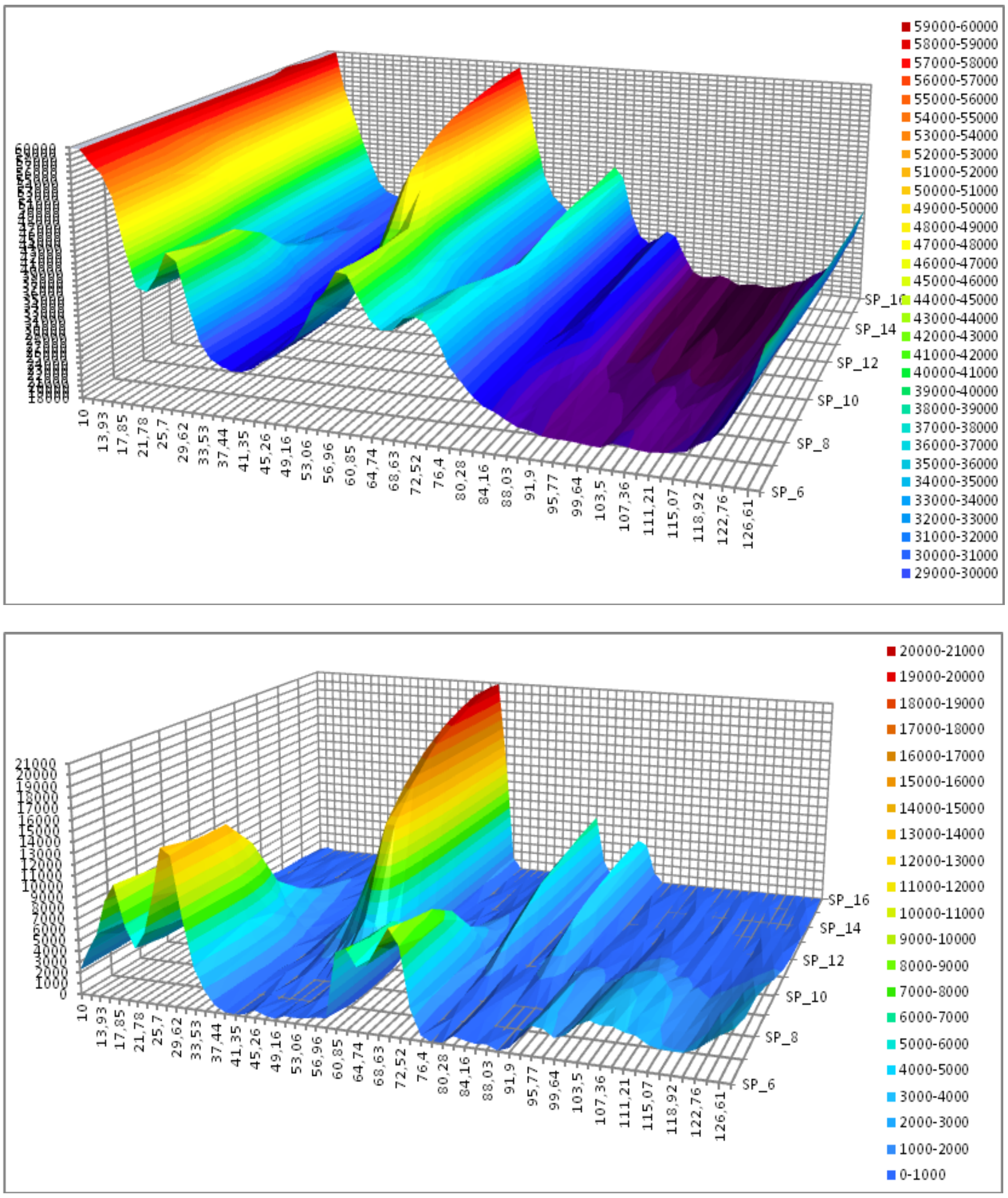

Fig.6. Kinetics of o-benzoquinone consumption and accumulation of gallic acid in the system during the photo-induced reduction of the product. (a) Reference THz Raman spectra, (b) the same spectra after the processing (deduction of the “minimal surface”)

The molecular probe, gallic acid, was chosen as a ligand of the catecholate type on the assumption that ortho-benzoquinone is formed on the surface of nanosize titanium dioxide particles [5]. A quantum-chemical simulation of the formation of the donor-acceptor complex of didentate (cyclic) type (“gallic acid-$TiO_2$”) was carried out under the GASSIAN environment. Calculations show that the formation of a ring of conjugation with the $TiO_2$ surface, in addition to the benzene ring, is probable, with the appearance of bands at 1500 $cm^{-1}$ and 878 $cm^{-1}$ in the Raman spectrum. Indeed, rather strong bands are observed at 1503$cm^{-1}$ and 872 $cm^{-1}$ in the experimentally obtained spectra. When quinone is reduced to the initial phenolic compound, these bands are preserved, which shows that the didentate (cyclic) bond between the ligand and the gallic acid dimer persists on the $TiO_2$ surface. The reduction of the phenolic compound to 3,4,5-trihydroxybenzoic (gallic) acid as a result of the photo-induced protonation indicates that the assumption that just ortho-benzoquinone is formed on the surface of titanium dioxide as a result of oxidation is justified (see fig.7).

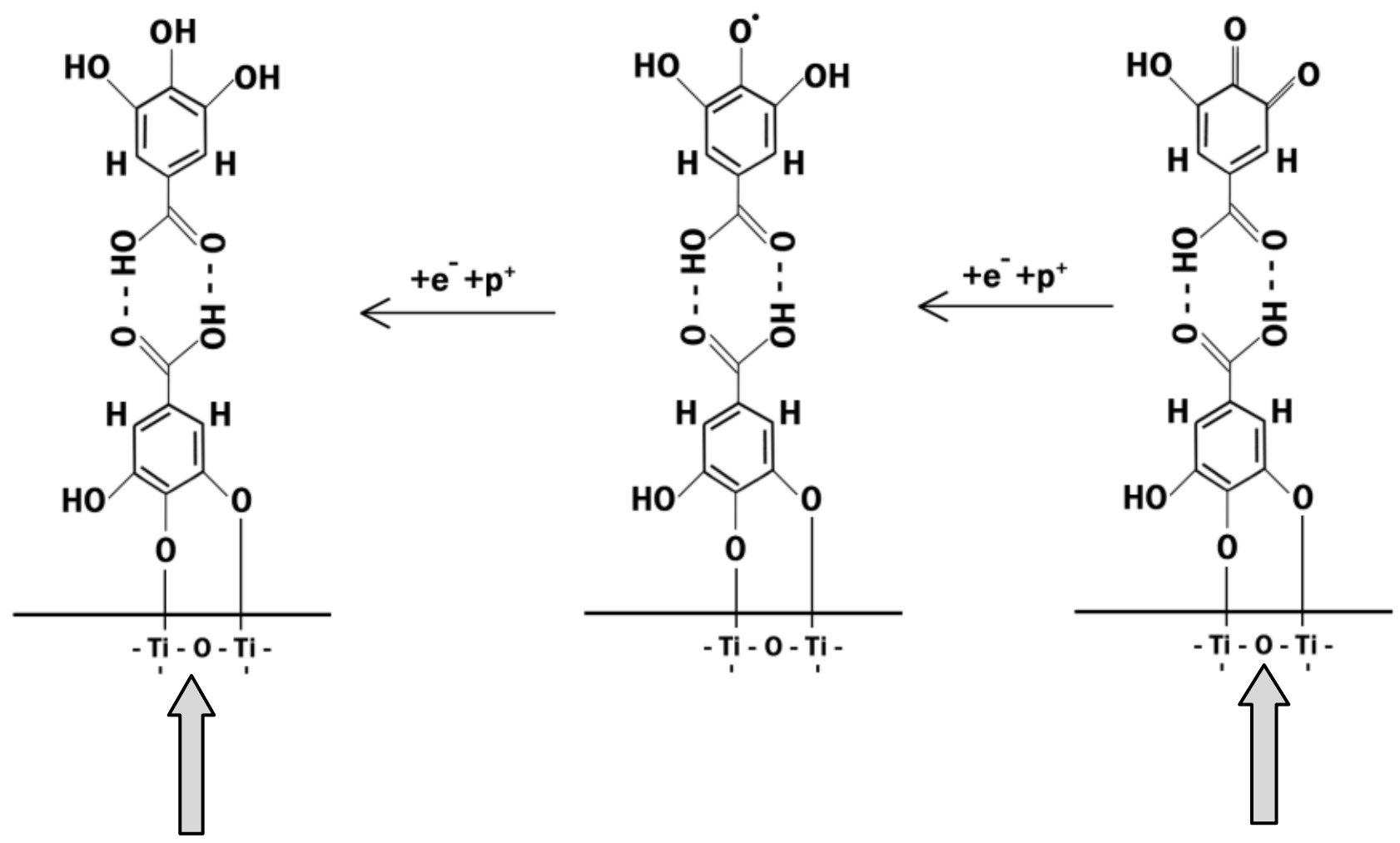

dimer Gallic Acid on $TiO_2$ dimer ortho-Benzoquinone on $TiO_2$

Fig.7. Diagram of the photo-induced reduction of ortho-benzoquinone to the gallic acid dimer.

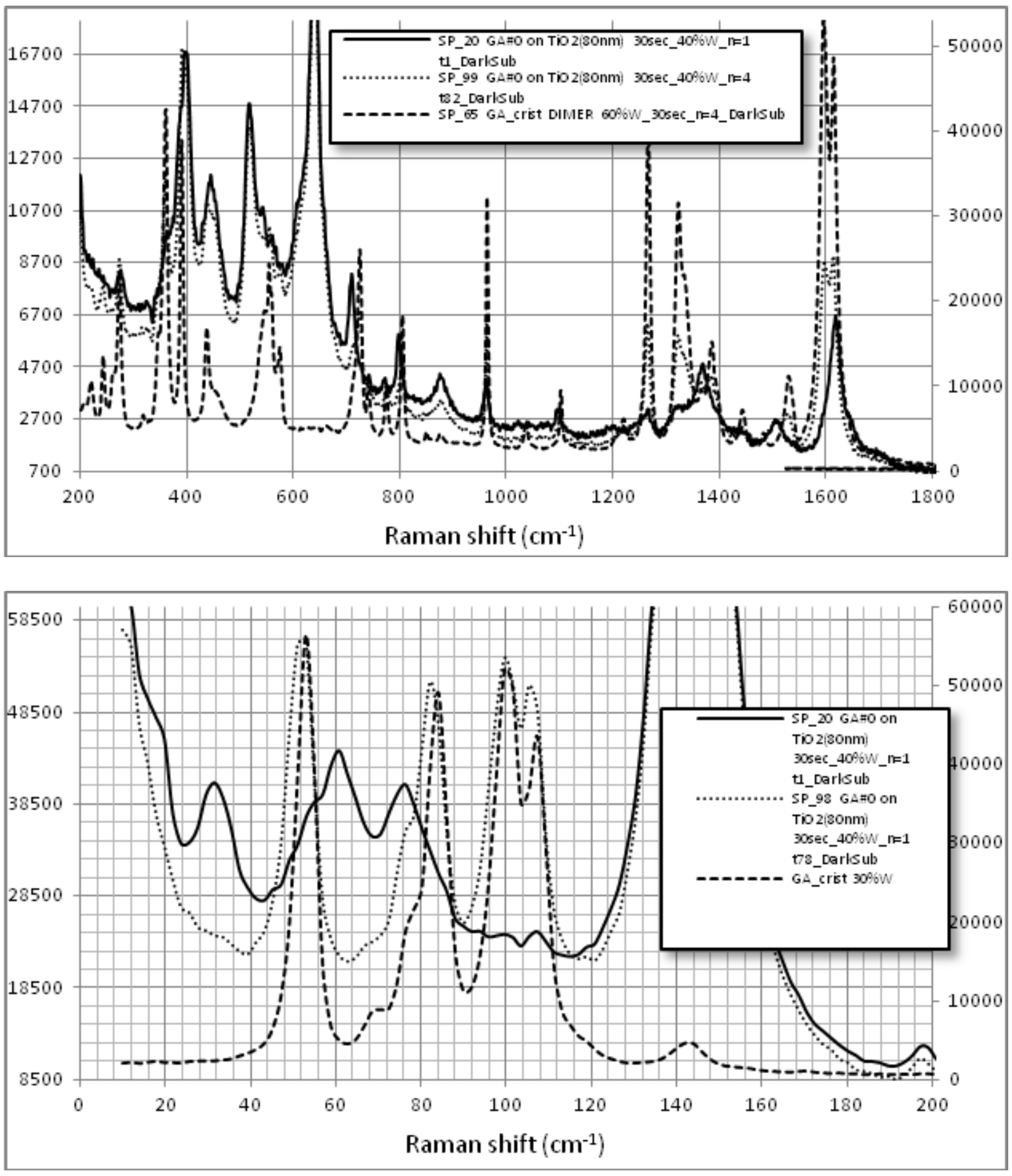

Fig.8. THz Raman spectra of the ortho-benzoquinone dimer on the surface of titanium dioxide and of the product of its reduction, gallic acid dimer adsorbed on $TiO_2$. The spectrum of gallic acid in the free state is given for reference.

Conclusions:

1. Raman and THz-Raman spectra of o-benzoquinones being formed upon oxidation of the gallic acid dimer on the surface of nanosize $TiO_2$ particles were for the first time obtained in experiments. Data on the kinetics of labile and stable products of the process were obtained.

2. A method was suggested for enhancing the contrast of spectral lines in kinetic studies of the process in which phenols are oxidized and quinones are reduced.

3. A kinetic (“reflective”) method was proposed for identification of individual substances in a complex multi-component system, which are formed and consumed during the oxidation and reduction reactions under heterogeneous conditions.

4. The photo-induced restoration of the THz-Raman spectrum profile of gallic acid was discovered, which makes it possible to correlate the SERS spectra available from experiments with spectral databases.